\documentclass{ws-procs975x65}%
\usepackage{amsmath}
\usepackage{amsfonts}
\usepackage{amssymb}
\usepackage{graphicx}%
\begin{document}

\title{Covariant Renormalizable Gravity Theories\\ on (Non) Commutative Tangent Bundles}

\author{Sergiu I. Vacaru}

\address{Alexandru Ioan Cuza University,
  14 Alexandru Lapu\c sneanu street,  Corpus R, UAIC, office 323;
Ia\c si,\  700057,\  Romania
\email{E--mails:\ sergiu.vacaru@uaic.ro, Sergiu.Vacaru@gmail.com}}

\begin{abstract}
The  field equations in modified gravity   theories possess an important decoupling property with respect to certain classes of nonholonomic frames. This allows us to construct  generic off--diagonal solutions depending on all spacetime coordinates via generating and integration functions containing (un--)broken symmetry parameters. Some corresponding analogous  models have a nice ultraviolet behavior and seem to be  (super) renormalizable in a sense of covariant  modifications of Ho\v{r}ava--Lifshits (HL) and ghost free gravity.  The apparent noncommutativity and breaking of Lorentz invariance by quantum effects can be encoded into geometric objects and basic equations  on noncommutative tangent Lorentz. The constructions can be extended to include conjectured covariant reonormalizable models with  effective Einstein fields with (non)commutative variables.
\end{abstract}

\bigskip\bodymatter

We study four equivalent models determined by actions of type
\begin{equation}
S=\frac{1}{16\pi }\int \delta u^{4}\sqrt{|\mathbf{g}_{\alpha \beta }|}\
\mathcal{L},\mbox{ for }\mathcal{L=}\ ^{[i]}\mathcal{L}, \label{act}
\end{equation}%
 where  $\ ^{[1]}\mathcal{L} = \widehat{f}(\widehat{R})+\ ^{m}L,\ ^{[2]}\mathcal{L}= \widetilde{R}+\widetilde{L}, \ ^{[3]}\mathcal{L} = R+L(T^{\mu \nu },R_{\mu \nu }),\ ^{[4]}\mathcal{L}=f(\breve{R})+\ ^{m}L$  are considered for the same  metric  $\mathbf{g=\{g}_{\mu \nu }\}$ on a four dimensional (4d) Lorentz manifold $\mathbf{V}$  but for different connections and respective scalar curvatures  and/ or $f$--modified Lagrange densities \cite{odints1} ; $\ ^{m}L$ and $L$ are  respectively the  Lagrange densities of matter fields and effective matter. The spacetime  axiomatic can be extended\cite{vfinsl} for Einstein -- Finsler like models on tangent Lorentz bundle $T\mathbf{V}$.

 Theories with $\ ^{[1]}\mathcal{L}$ are useful for constructing  generic off--diagonal solutions in general relativity, GR, and modifications.  Geometric models with $\ ^{[2]}\mathcal{L}$ can  be considered as "bridges" between generating functions determining certain classes of generic off--diagonal Einstein manifolds and modified gravity theories. The effective Lagrange density $\ ^{[3]}\mathcal{L}$ is similar to that introduced for covariant renormalization and HL  gravity  \cite{odints1}. The models with $\ ^{[4]}\mathcal{L}$ transfers us into the field of "standard" modified theories with scalar curvature determined by the Levi--Civita (LC) connection. In almost K\"{a}hler variables, theories of type $\ ^{[1]}\mathcal{L}$--$\ ^{[3]}\mathcal{L}$ can be quantized using non--perturbative methods for deformation quantization and A--brane quantization, or as perturbative gauge theories \cite{vquant}.

A class of (quantum type) modified noncommutative gravity theories are physically motivated by the Schr\"{o}dinger type uncertainty relations $\hat{u}^{\alpha }%
\hat{p}^{\beta }-\hat{p}^{\beta }\hat{u}^{\alpha }=i\hbar \widehat{\theta }^{\alpha \beta },$ where $\hbar $ is the Planck constant, $i^{2}=-1,$ and $%
\hat{u}^{\alpha }$ and $\hat{p}^{\beta }$ are, respectively, certain
coordinate and momentum type operators. Such geometric models are elaborated on "$\theta $--extensions" of tangent
Lorentz bundles $T\mathbf{V\rightarrow }$ $^{\theta }T\mathbf{V},$ (there
are used also co--tangent bundles $T^{\ast }\mathbf{V\rightarrow \ }$ $ ^{\theta }T^{\ast }\mathbf{V}$ etc). The extensions are determined by
noncommutative complex distributions stated by "generalized uncertainty" relations
 $u^{\alpha _{s}}u^{\beta _{s}}-u^{\beta _{s}}u^{\alpha _{s}}=i\theta ^{\alpha
_{s}\beta _{s}},$  
where the antisymmetric matrix $\theta =(\theta ^{\alpha _{s}\beta _{s}})$
can be taken with constant coefficients with respect to certain frame of
reference (for "small" $\theta $--deformations $\mathbf{V\rightarrow \
^{\theta }V}$). The label "$s$" is considered for three two dimensional (2-d)
"shells" $s=0,1,2$ and a conventional splitting on $^{\theta }T\mathbf{V,}$
$\dim (\ ^{\theta }T\mathbf{V})=4+2s=2+2+2+2=8$ with coordinates respectively parameterized $u^{\alpha
_{s}}=(x^{i_{s}},y^{a_{s}}).$
The uncertainty relations  and coordinate parameterizations  are adapted to Whitney sums (nonlinear connections, N--connections) $\ ^{s}\mathbf{N}:T\ ^{s}\mathbf{V}=h\mathbf{V}\oplus v\mathbf{V}\oplus \ ^{1}v\mathbf{V}\oplus \ ^{2}v\mathbf{V.}$ This prescribes a local fibred structure on $\mathbf{\ }%
^{\theta }\mathbf{V,}$ when  $\ ^{s}\mathbf{N}=N_{i_{s}}^{a_{s}}(\ ^{s}u,\theta
)dx^{i_{s}}\otimes \partial /\partial y^{a_{s}}$  defines N--adapted (co) bases  $\mathbf{e}%
_{\nu _{s}}=(\mathbf{e}_{i_{s}},e_{a_{s}})$ and  $\mathbf{e}^{\mu _{s}}=(e^{i_{s}},\mathbf{e}^{a_{s}})$;\
$\mathbf{e}_{i_{s}} =\frac{\partial }{\partial x^{i_{s}}}-\
N_{i_{s}}^{a_{s}}\frac{\partial }{\partial y^{a_{s}}},\ e_{a_{s}}=\frac{%
\partial }{\partial y^{a_{s}}};\  e^{i_{s}} = dx^{i_{s}},\mathbf{e}^{a_{s}}=dy^{a_{s}}+\
N_{i_{s}}^{a_{s}}dx^{i_{s}}.$

Any metric  $\ ^{s}\mathbf{g}$ on $\ ^{\theta }\mathbf{V}$  can be represented as a distinguished metric (d--metric)
{\small
\begin{eqnarray}
 \ ^{s}\mathbf{g} &=& =\{\mathbf{g}_{\alpha _{s}\beta _{s}}\} =\ g_{i_{s}j_{s}}(\ ^{s}u,\theta )\ e^{i_{s}}\otimes
e^{j_{s}}+\ g_{a_{s}b_{s}}(\ ^{s}u,\theta )\mathbf{e}^{a_{s}}\otimes \mathbf{%
e}^{b_{s}}  \label{dm1s} \\
&=&g_{ij}(x)\ e^{i}\otimes e^{j}+g_{ab}(u)\ \mathbf{e}^{a}\otimes \mathbf{e}%
^{b}+ g_{a_{1}b_{1}}(\ ^{1}u,\theta )\ \mathbf{e}^{a_{1}}\otimes \mathbf{e}%
^{b_{1}}+\ g_{a_{2}b_{2}}(\ ^{2}u,\theta )\mathbf{e}^{a_{2}}\otimes \mathbf{e%
}^{b_{2}}.  \notag
\end{eqnarray}%
}
 A self--consistent approach to such theories is based on the Groenewold--Moyal product (star
product, or $\star $--product).  We apply a geometric formalism  \cite{vasil}  modified
for nonholonomic distributions and connections with $\nabla \rightarrow
\mathbf{D}=\mathbf{\tilde{D},}$ or $=\widehat{\mathbf{D}},$ and almost K%
\"{a}hler variables determined  by data $(\mathbf{g,N})$. 
The symbol $\mathbf{\tilde{D}}$ is used for the Cartan connection
completely determined by regular effective Lagrange generating function $%
\mathcal{L}(x,y),$  and a canonical almost K\"{a}hler space with $\mathbf{g}=\mathbf{\tilde{g},}$ $\mathbf{N=%
\tilde{N}}$ and $\mathbf{J=\tilde{J}}$  defined by $\mathcal{L},$
when $\mathbf{\tilde{\theta}(\cdot ,\cdot )}:=\mathbf{\tilde{g}}\left(
\mathbf{\tilde{J}\cdot ,\cdot }\right) .$ 

It is possible to define  $s$--shell associative star operators $\ ^{s}\tilde{\star}$\ if $\mathbf{D%
}_{\mu _{s}}=(\mathbf{D}_{i_{s}},\mathbf{D}_{a_{s}}),$ for $\mathbf{D}=\mathbf{\tilde{D}}$ (or $=$ $\widehat{\mathbf{D}}),$
 $\alpha \ (\ ^{s}\tilde{\star})\beta =\sum\limits_{k}\frac{\ell ^{k}}{k!}%
\mathbf{\theta }^{a_{1}b_{1}}...\mathbf{\theta }^{a_{k}b_{k}}(\mathbf{D}%
_{a_{1}}\ldots \mathbf{D}_{a_{k}})\cdot (\mathbf{D}_{b_{1}}\ldots \mathbf{D}%
_{b_{k}})$.
 We derive a N--adapted local frame structure $\mathbf{\tilde{e}}_{\alpha _{s}}=(\mathbf{e}_{i_{s}},%
\mathbf{\tilde{e}}_{a_{s}})$ which can be constructed by noncommutative
deformations of $\mathbf{e}_{\alpha },$\
 $ \mathbf{\tilde{e}}_{\alpha _{s}\ }^{\ \underline{\alpha }_{s}} = \mathbf{e}%
_{\alpha _{s}\ }^{\ \underline{\alpha }_{s}}+i\theta ^{\gamma _{s}\beta _{s}}%
\mathbf{e}_{\alpha _{s}\ \gamma _{s}\beta _{s}}^{\ \underline{\alpha }%
_{s}}+\theta ^{\gamma _{s}\beta _{s}}\theta ^{\tau _{s}\mu _{s}}\mathbf{e}%
_{\alpha _{s}\ \gamma _{s}\tau _{s}\mu _{s}}^{\ \underline{\alpha }_{s}}+%
\mathcal{O}(\theta ^{3}),\  
\mathbf{\tilde{e}}_{\ \star \underline{\alpha }_{s}}^{\alpha _{s}} =%
\mathbf{e}_{\alpha _{s}\ }^{\ \underline{\alpha }_{s}}+i\theta ^{\gamma
_{s}\beta _{s}}\mathbf{e}_{\ \underline{\alpha }_{s}\gamma _{s}\beta
_{s}}^{\alpha _{s}}+\theta ^{\gamma _{s}\beta _{s}}\theta ^{\tau _{s}\mu
_{s}}\mathbf{e}_{\ \underline{\alpha }_{s}\gamma _{s}\beta _{s}\tau _{s}\mu
_{s}}^{\alpha _{s}}+\mathcal{O}(\theta ^{3})$,  
where $\mathbf{\tilde{e}}_{\ \star \underline{\alpha }%
_{s}}^{\alpha _{s}}\star \mathbf{\tilde{e}}_{\alpha _{s}\ }^{\ \underline{%
\beta }_{s}}\ =\delta _{\underline{\alpha }_{s}}^{\ \underline{\beta }_{s}},$
 $\delta _{\underline{\alpha }}^{\ \underline{\beta }}$ is the
Kronecker tensor. The values $\mathbf{e}_{\alpha _{s}\ \gamma _{s}\beta
_{s}}^{\ \underline{\alpha }_{s}}$ and $\mathbf{e}_{\alpha _{s}\ \gamma
_{s}\tau _{s}\mu _{s}}^{\ \underline{\alpha }_{s}}$ can be written in terms
of $\mathbf{e}_{\alpha _{s}\ }^{\ \underline{\alpha }_{s}},\theta ^{\gamma
_{s}\beta _{s}}$ and the spin distinguished connection corresponding to $\
^{s}\mathbf{\tilde{D},}$ or $\ ^{s}\widehat{\mathbf{D}}$.   We can parameterize
the noncommutative and nonholonomic transforms when $\ ^{s}\mathbf{g}%
_{\alpha _{s}\beta _{s}}(\ ^{s}u,\theta )$ (\ref{dm1s}) is with real
coefficients which for $s=1,2$ depend only on even powers of $\theta ,\
g_{i}(u) = \grave{g}_{i}(x^{k}),\ h_{a}=\grave{h}_{a}(u),\ h_{a_{s}}=\grave{%
h}_{a_{s}}(u)+\ ^{2}h_{a_{s}}(u)\theta ^{2}+\mathcal{O}(\theta ^{4}),\
N_{i_{s}}^{a_{s}}(\ ^{s}u,\theta ) =\grave{N}_{i_{s}}^{a_{s}}(\ ^{s}u)+\
^{2}N_{i_{s}}^{a_{s}}(\ ^{s}u)\theta ^{2}+\mathcal{O}(\theta ^{4})$.
  For simplicity, we shall not write $\theta $ in
explicit form if that will not result in ambiguities for any notation of type $N_{i_{s}}^{a_{s}}(\ ^{s}u,\theta
),N_{i_{s}}^{a_{s}}(\theta ),$ or $N_{i_{s}}^{a_{s}}(\ ^{s}u).$

Applying a N--adapted variational calculus for $\ ^{[1]}\mathcal{L}$ on $\mathbf{V}$, we derive the equations of motion for  4--d modified massive gravity \cite{odints1,vexsolbranes}
\ $(\partial _{\widehat{R}}\widehat{f})\widehat{\mathbf{R}}_{\alpha \beta }-%
\frac{1}{2}\widehat{f}(\widehat{R})\mathbf{g}_{\alpha \beta }+\mathring{\mu}%
^{2}\mathbf{X}_{\alpha \beta }=M_{Pl}^{-2}\mathbf{T}_{\alpha \beta },$ 
where $M_{Pl}$ is the Plank mass, $\widehat{\mathbf{R}}_{\mu \nu }$ is the
Einstein tensor for a pseudo--Riemannian metric $\mathbf{g}_{\mu \nu }$ and $%
\widehat{\mathbf{D}},$ $\mathbf{T}_{\mu \nu }$ is the standard
energy--momentum tensor for matter.  For simplicity, we  consider sources which (using frame transforms) can be parameterized with respect to
N--adapted frames  in the form
 $\Upsilon _{\beta }^{\alpha }=\frac{1}{M_{Pl}^{2}(\partial _{\widehat{R}}%
\widehat{f})}\mathbf{T}_{\beta }^{\alpha }=\frac{1}{M_{Pl}^{2}(\partial _{\widehat{R}}\widehat{f})} \
^{m}T\ \delta _{\beta }^{\alpha }=\hat{\Upsilon}\ \delta _{\beta }^{\alpha }$,
 for a constant value $\ \hat{\Upsilon}:=M_{Pl}^{-2}(\partial _{\widehat{R}}%
\widehat{f})^{-1}\ ^{m}T$. Under general assumptions, the effective source can be parameterized
 $\mathbf{\Upsilon }_{~\delta _{s}}^{\beta _{s}}=(\ ^{s}\hat{\Upsilon}+\ \ ^{s}%
\mathring{\Upsilon})\mathbf{\delta }_{~\delta _{s}}^{\beta _{s}}$.
 The sources with $s=1$ and $2$ are considered as certain effective ones on
"fibers" with noncommutative variables.  On $ ^{\theta }T\mathbf{V,}$ the modified gravitational field equations  are 
\begin{equation}
\ ^{s}\widehat{\mathbf{R}}_{\ \beta _{s}\delta _{s}}-\frac{1}{2}\mathbf{g}%
_{\beta _{s}\delta _{s}}\ ^{sc}\widehat{R}=\mathbf{\Upsilon }_{\beta
_{s}\delta _{s}},\   
\widehat{L}_{a_{s}j_{s}}^{c_{s}}=e_{a_{s}}(N_{j_{s}}^{c_{s}}),\ \widehat{C}%
_{j_{s}b_{s}}^{i_{s}}=0,\ \Omega _{\ j_{s}i_{s}}^{a_{s}}=0.  \label{cdeinst} 
\end{equation}%

We generate solutions of (\ref{cdeinst}) by any 
{\small
\begin{eqnarray}
ds_{K}^{2} &=&\epsilon _{i}e^{\psi (x^{k},\mathring{\mu})}(dx^{i})^{2}+\frac{%
\ \tilde{\Phi} ^{2}}{4\widetilde{\Lambda }}\left[ dy^{3}+(\partial _{i}\ n)dx^{i}%
\right] ^{2}+\ \frac{(\partial _{4}\ \tilde{\Phi} )^{2}}{\ \widetilde{\Lambda }\
\Phi ^{2}}\left[ dy^{4}+(\partial _{i}\ \check{A})dx^{i}\right] ^{2}  \notag
\\
&&+\frac{\ ^{1}\tilde{\Phi}^{2}}{4\ ^{1}\widetilde{\Lambda }}\left[
dy^{5}+(\partial _{\tau }\ ^{1}n)du^{\tau }\right] ^{2}+\ \frac{(\partial
_{6}\ ^{1}\tilde{\Phi})^{2}}{\ ^{1}\widetilde{\Lambda }\ ^{1}\tilde{\Phi}^{2}%
}\left[ dy^{6}+(\partial _{\tau }\ ^{1}\check{A})du^{\tau }\right] ^{2}
\label{qellcs} \\
&&+\frac{\ ^{2}\tilde{\Phi}^{2}}{4\ ^{2}\widetilde{\Lambda }}\left[
dy^{7}+(\partial _{\tau _{1}}\ ^{2}n)du^{\tau _{1}}\right] ^{2}+\ \frac{%
(\partial _{8}\ ^{2}\tilde{\Phi})^{2}}{\ ^{2}\widetilde{\Lambda }\ ^{2}%
\tilde{\Phi}^{2}}\left[ dy^{8}+(\partial _{\tau _{1}}\ ^{2}\check{A}%
)du^{\tau _{1}}\right] ^{2}.  \notag
\end{eqnarray}%
}

The Lagrangian densities (\ref{act})  include $z=3$ theories which allows us to generate ultra--violet
power counting renormalizable $3+1$ and/or $2+2$ quantum models. We can take
more general sources and generating functions instead of $~^{\diamond }L$
and write {\small
\begin{equation}
-\ _{\hat{n}}^{\diamond }L=\hat{\alpha}\{(T^{\mu \nu }\ ^{\diamond }\nabla
_{\mu }\ ^{\diamond }\nabla _{\nu }+\hat{\gamma}T_{\alpha }^{\alpha }\
^{\diamond }\nabla ^{\beta }\ ^{\diamond }\nabla _{\beta })^{\hat{n}%
}(T^{\alpha \beta }\ ^{\diamond }R_{\alpha \beta }+\hat{\beta}T_{\alpha
}^{\alpha }\ ^{\diamond }R_{\beta }^{\beta })\}^{2},  \label{auxa}
\end{equation}%
} where $\hat{n}$ and $\hat{\gamma}$ are constants and $\ ^{\diamond }\nabla
_{\mu }$ is obtained via nonholonomic constraints of any $\ ^{\diamond }%
\mathbf{D}_{\mu }$ which is metric compatible and completely determined by a
metric structure.  We use $\Upsilon =~-~_{\hat{n}}^{\diamond }L$  which results in a different class of generating
functions $~^{\hat{n}}\widetilde{\phi }$ from $\frac{\partial _{4}(~^{\hat{n}%
}\widetilde{\phi })\partial _{4}h_{3}}{2h_{3}h_{4}}=-~_{\hat{n}}^{\diamond
}L.$ A new class of off--diagonal solutions (\ref{qellcs}) can be generated
for data $\widetilde{\phi }\rightarrow ~^{\hat{n}}\widetilde{\phi },\phi
\rightarrow ~^{\hat{n}}\phi $ and $~^{\diamond }\Upsilon \rightarrow -~_{%
\hat{n}}^{\diamond }L.$ Twisted (non) commutative extensions to higher
"velocity/ momentum" type fibers can be performed in a standard manner
as we considered in the previous section.  We can apply
the analysis provided in \cite{odints1,vquant} which states that the
analogous $\ ^{[3]}\mathcal{L}$ and $\ ^{[4]}\mathcal{L}$ models derived for
(\ref{auxa}) are renormalizable if $\hat{n}=1$ and super--renormalizable for
$\hat{n}=2.$ Such properties can be preserved for theories generalized on
tangent Lorentz bundles.

{\bf Acknowledgments:} Authors's research is partially supported by the Program IDEI, PN-II-ID-PCE-2011-3-0256.


\begin{thebibliography}{9}                                                                                                %
\bibitem{odints1} S. Nojiri, S. D. Odintsov, \textsl{Phys. Lett.} \textbf{B691},  60 (2010);  P. Ho\v{r}ava, Phys. Rev. D 79 (2009) 084009

\bibitem{vfinsl} N. Mavromatos et all, \textsl{Eur. Phys. J.} \textbf{C72}, 1956 (2012); P. Stavrinos, S. Vacaru, \textsl{Class. Quant. Grav.} \textbf{30}, 055012 (2013);\ S. Vacaru, \textsl{Int. J. Mod. Phys.} \textbf{D 21},  1250072 (2012); \textsl{Class. Quant. Grav.} \textbf{28},  215991 (2011)

\bibitem{vquant} S. Vacaru, \textsl{Eur. Phys. J.} \textbf{C 73},  2287 (2013);\ \textsl{EPL} \textbf{96},  5001 (2011); \textsl{Int. J. Geom. Meth. Mod. Phys.} \textbf{7}, 713 (2010);\  \textsl{J. Geom. Phys.} \textbf{60},  1289 (2010)

\bibitem{vasil} D. V. Vassilevich, \textsl{Class. Quant. Grav.} \textbf{26},  145010 (2009);\  \textbf{27},  095020 (2010);  Amelino-Camelia et all, \textsl{Class. Quant. Grav.} \textbf{29},  075007 (2012); S. Vacaru, \textsl{Class. Quant. Grav.} \textbf{27},  105003 (2010); \textsl{J. Math. Phys.} \textbf{50}, 073503 (2009) 

\bibitem{vexsolbranes} N. Mavromatos, S. Sarkar, A. Vergou, \textsl{Phys. Lett.} \textbf{B696},  300 (2011); S. Vacaru, \textsl{Gen. Relat. Grav.} \textbf{44},  1015 (2012);\  \textsl{Int. J. Geom. Meth. Mod. Phys.} \textbf{4},  1285 (2007)
\end{thebibliography}
\end{document}